\documentclass[article]{IEEEtran}
\IEEEoverridecommandlockouts

\usepackage{cite}
\usepackage{graphics} 
\usepackage{epsfig} 
\usepackage{eurosym}
\usepackage{subcaption}
\usepackage{soul,color}
\usepackage{acronym}
\usepackage{float}
\usepackage{epstopdf}
\usepackage{amsmath}
\usepackage{rotating}
\usepackage{array}
\usepackage{booktabs}
\usepackage{dcolumn}
\usepackage{multicol, blindtext}
\usepackage{multirow}
\usepackage{makecell,lineno}
\usepackage{url}
\usepackage{amsmath}
\usepackage{amssymb}

\makeatletter
\def\algbackskip{\hskip-\ALG@thistlm}
\makeatother

\usepackage{algorithm}
\usepackage[noend]{algpseudocode}
\usepackage{courier}

\acrodef{VWBump}[VWBump]{Variable Width Bump}
\acrodef{EMG}[EMG]{Electromyographic}
\acrodef{DPI}[DPI]{Diff-Pair Integrator}
\acrodef{TCA}[TCA]{Transconductance Amplifier}

\begin{document}

\title{Digital Multiplier-less Event-Driven Spiking Neural Network Architecture for Learning a Context-Dependent Task }

\author{Hajar~Asgari,
        Babak~Mazloom-Nezhad~Maybodi,
        Raphaela~Kreiser,
        and~Yulia~Sandamirskaya
\thanks{H. Asgari is a visiting Ph.D. student at Institute of Neuroinformatics (INI), Zurich, Switzerland from Shahid Beheshti University of Tehran, Iran (e-mail: h\_asgari@sbu.ac.ir). }
\thanks{ B. Mazloom-Nezhad Maybodi is with the Department of Electrical Engineering, Shahid Beheshti University, Tehran, Iran (e-mail: b-mazloom@sbu.ac.ir).}
\thanks{R. Kreiser and Y. Sandamirskaya are with the Institute of Neuroinformatics (INI), University of Zurich and ETH Zurich, Zurich, Switzerland (e-mail: rakrei, ysandamirskaya@ini.uzh.ch)}}

\markboth{IEEE Transactions on Circuits and Systems I: Regular Papers (TCAS-I)}%
{Shell \MakeLowercase{\textit{et al.}}: Bare Demo of IEEEtran.cls for IEEE Journals}

\maketitle

\begin{abstract}
Neuromorphic engineers aim to develop event-based spiking neural networks (SNNs) in hardware. These SNNs closer resemble dynamics of biological neurons  than todays' artificial neural networks and achieve higher efficiency thanks  to the event-based, asynchronous nature of processing. Learning in SNNs is more challenging, however. Since conventional supervised learning methods cannot be ported on SNNs due to the non-differentiable event-based nature of their activation, learning in SNNs is currently an active  research topic. Reinforcement learning (RL) is particularly promising method for neuromorphic implementation, especially in the field of autonomous agents' control, and is in focus of this work. In particular, in this paper we propose a new digital multiplier-less hardware implementation of an SNN. 
We show how this network can learn stimulus-response associations in a context-dependent task through a RL mechanism. The task is inspired by biological experiments used to study RL in animals. The architecture is described using the standard digital design flow and uses power- and space-efficient cores.
We implement the behavioral experiments using a robot, to show that learning in hardware also works in a closed sensorimotor loop.

\end{abstract}


\IEEEpeerreviewmaketitle
\section{Introduction} 
By the end of the Moore's law, there is a need for new computing architectures that respect the severe cost and power reduction constraints. In contrast to the von Neumann processing architecture, the biological brain includes billions of neurons and thousands of synapses per neuron with high fan-out event-driven communication to attain exceptional energy efficiency~\cite{And2015, Mer2014, Charlotte2018}.
During the past few years, SNNs have received considerable attention in the artificial neural network community due to their performance-resource trade-off. On the other hand, the neurosciece community is also interested in their behavioral resemblance to biological neurons\cite{Izh2003, Izh2004}.
In spiked-based platforms parallel computational units (neurons) produce events that model neurons' spikes. Spikes produced by source neurons are transmitted to one or more destination synapses that alter these incoming signals with different gain factors and temporal filtering, and convey them to the post-synaptic neuron. 
In these event-driven SNNs, information can be encoded in the timing of these events, their firing rates, or stable activity patterns in large neuronal populations~\cite{Saber2018,SchonerSpencer2015}.\\
Accordingly, several research groups around the world recently have proposed approaches for building reconfigurable spiked-based neural processing architectures~\cite{Benj2015, Ning2015, Ind2007, Charlotte2018, Saber2018, Davi2018, Jok2017, Hay2015, Sol2012, Nou2017, Fil2015}.
The Field-Programmable Gate Arrays (FPGAs) have advantageous features for prototyping neuromorphic systems, such as flexibility, reliability, and reconfigurability~\cite{ Darwin2017,Lammie2018,Edris2019}. Lately, such reconfigurable digital platforms have been utilized to realize models of the biological nervous system~\cite{Sol2012, Hay2015,  Nou2017}.
Most of the recent spiking neural network models on hardware have addressed pattern classification and vision applications and demonstrated outstanding improvements in energy efficiency  compare to software computations~\cite{Charlotte2018, Ning2015, Edris2019}. However, besides needs for image classification, autonomous systems such as mobile robots are reported as a potential field for neuromorphic applications~\cite{Elisa2014,MildeEtAl2017}. 
Motivated by these findings this paper proposes a digital circuit implementation of an event-driven spiking neural network. We use this model to demonstrate learning a context-dependent task through reinforcement learning~\cite{flor2014}, suitable for realization of online learning on an autonomous agent. Such context-dependent tasks are used in biology to explore reinforcement learning in situations, where different state-action value functions need to be learned, depending on the context.
RL  is used to model animal behavior in such a task. The proposed digital spiking network model realizes such reinforcement learning in hardware. We show that the system  is able to learn the link between an item and a context while developing independence for the place, as observed in computational models of animal experiments~\cite{Komo2009, flor2014}.  
Reinforcement learning is a method of learning how to make a correct decision in an autonomous agent~\cite{Sutton2018}. In RL, the autonomous agent is trained by interacting with an environment and receiving rewards based on its experience. In pursuit of learning, the agent  senses the environment, selects an action, and updates the internal representation of the value of the state, the action, or the state-action pair based on an intermittent rewarding signal. Eventually, the agent learns to maximize the cumulative reward in the long run.  Thus,  RL is a biologically  inspired algorithm, which can provide true autonomy in  intelligent systems~\cite{Anvesha2018}. 

Modelling SNN for RL includes high accuracy calculations of the membrane potential and, consequently, for implementing these networks, high precision leaky integrate-and-fire (LIF) neuron and spike-timing-dependent plasticity (STDP) based synapse modules are needed. 
We implement the network on an FPGA chip and compare the hardware model results during learning with the computational model of rat behavior and the animal experiment~\cite{flor2014, Komo2009}.

The main contributions of the manuscript are the following:

\begin{itemize}
\item Proposing a new multiplier-less event-driven SNN architecture for a context-dependent task through reinforcement learning algorithm.
\item Presenting a simple high precision model for synaptic strength based STDP.
\item Introducing a digital multiplier-less model for LIF neuron based on state machine diagram concept.
\item Suggesting a new verification demo with a robotic experiment for reinforcement learning mechanism in neuromorphic applications.
\end{itemize}
The manuscript starts with a description of the context-dependent task, the SNN model, and the learning mechanism in Section II. After that, we introduce the proposed hardware implementation methods for neurons, synapses, and the overall system architecture in Section III. Then, in Section IV implementation results and verification reports by robotic experiment are presented. Also in section IV, we provide a discussion on the network and previous studies. Finally, section V concludes the paper.

\section{Background of Biological Model}
\subsection{Description of The Context Dependent Task}
\begin{figure}[h]
	\centering
	\includegraphics[width=0.7\columnwidth]{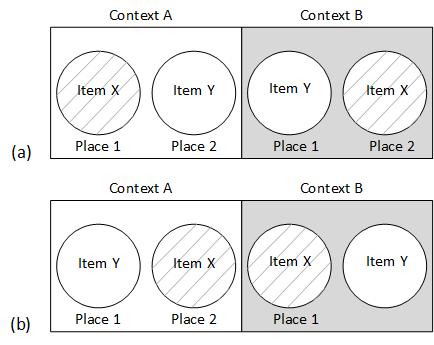}
	\caption{Schematics of the stimuli combinations, used in the animal experiments on context-derapendent reinforcement learning: In context A item X contains a buried reward and in context B item Y contains a buried reward. (a)~Triplets A1X, A2Y, B1X, B2Y. (b)~Triplets A1Y, A2X, B1Y, B2X } 
	\label{triplets}
\end{figure}

The context-dependent task was designed to examine the hippocampal principal neurons behavior during training which of two items is rewarded depending on the environment properties~\cite{Komo2009}. In the original experiment, there were two boxes as contexts A and B with a shared entrance which separates the boxes. Contexts had different kinds of wallpaper and floors. Before starting each trial, the rats were able to move between the two boxes. By the beginning of each trial boxes' entrance was closed and rats were isolated in one of them.
Each context has two positions  (positions 1 and 2). In each trial, two pots as Item X and Y were randomly located in two different positions. These pots differed in terms of filling material and only one of them contained a reward. In context A, Item X contained a reward whereas in context B, item Y contained a reward. In the model for this experiment, there are 4 spatial locations A1, A2, B1, B2. For example, A1 refers to position 1 in context A.
Due to the position of items X and Y and the contexts, eight different triplets appear. For example, triplet A1X means that item X locates in position 1 in context A. A group of these triplets is rewarded and another group is not. The rewarded group includes A1X, A2X, B1Y, and B2Y and the non-rewarded group contains A1Y, A2Y, B1X, and B2X.
After training, when the rat starts searching for the rewarded group triplets, it digs into the pots immediately; which means the rat gets closer to the pot and gets the hidden reward inside the pot. However, starting from the non-rewarded triplets most of the time causes a movement to another place in the same context. All combinations of contexts, items, and places in an experiment are shown in Fig.~\ref{triplets}. In Fig.~\ref{triplets}(a) A1X, A2Y, B1Y, and B2X are shown, while Fig.~\ref{triplets}(b) shows A1Y, A2X, B1X, and B2Y.

\subsection {Spiking Neural Network Model}
\begin{figure}[h]
	\centering
	\includegraphics[width=0.7\columnwidth]{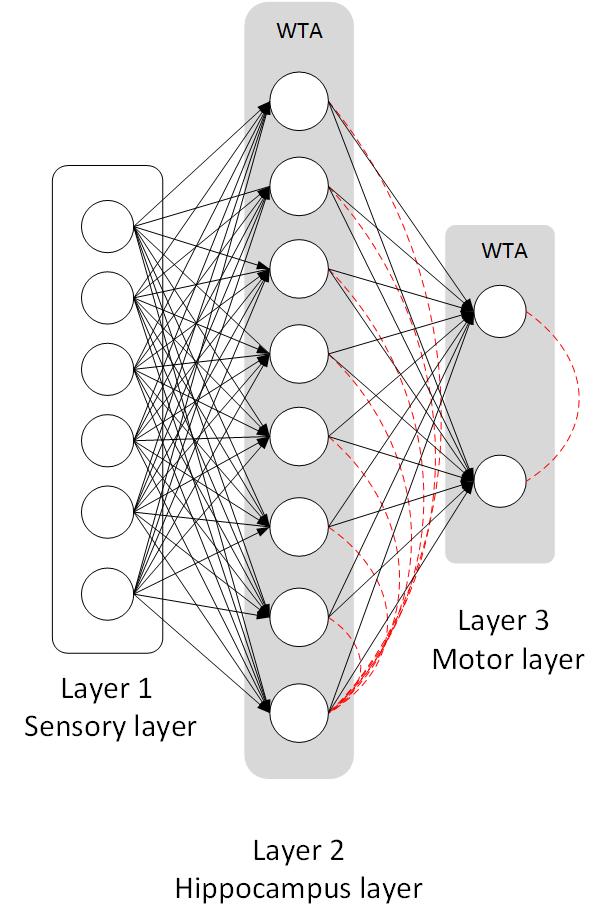}
	\caption{The spiking network, used to model the Hippocampal network involved in reinforcement learning in the context-dependent task in animals: an Input (sensory) layer with six neurons, a Hidden (Hippocampus) layer with eight neurons, and an output (motor) layer with two neurons. Excitatory all-to-all synaptic weights connect all neurons from input layer to hidden layer and from hidden layer to output layer (black lines). There are all to all inhibitory synaptic weights except self-connections  within neurons in the hidden and output layers (red lines).} 
	\label{SNN_model}
\end{figure}

\begin{figure}[h]
	\centering
	\includegraphics[width=0.8\columnwidth]{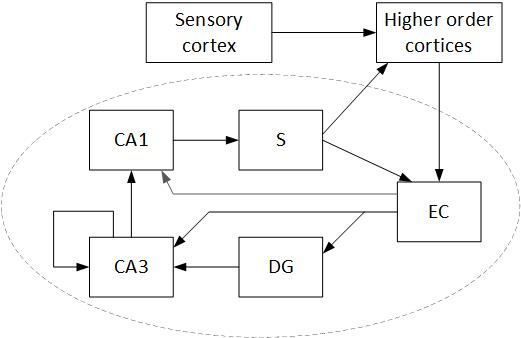}
	\caption{Anatomy of the hippocampal formation, involved in the reinforcement learning task: CA-cornu ammonis, DG-dentate
		gyrus, S-subiculum, EC-entorhinal cortex~\cite{Arthur2006}. } 
	\label{HippoAnatomy}
\end{figure}

The spiking neural network model for learning the context-dependent task is depicted in Fig.~\ref{SNN_model}. This network models CA1 region in Hippocampus which includes an Input (sensory) layer, a hidden (Hippocampal) layer and an output (motor) layer.
Fig.~\ref{HippoAnatomy} shows that the CA1 region in Hippocampus receives its motivation input from entorhinal cortex and sends its output back to the entorhinal cortex via subiculum or directly~\cite{Arthur2006}. Thus, the sensory layer of the SNN is related to a group of neurons in CA1 that receives motivation input from entorhinal cortex neurons. The motor layer in Fig.~\ref{SNN_model} corresponds to neurons in layer CA1 that are connected to entorhinal cortex neurons; Finally, the hippocampal layer models a few neurons in the CA1 region~\cite{flor2014}.
In this model, the input layer includes six neurons and provides context-place information. Adaptive unidirectional excitatory weights connect all these input neurons to all hippocampal neurons. The eight hippocampal neurons in the hidden layer have inhibitory connections among themselves, but they don't have self-inhibition.
A number of plastic excitatory synapses connect all hippocampal neurons to the output layer neurons. Similar to the hippocampus layer, neurons in the output layer have inhibitory connections within themselves. Thus, in this network sensory and output layers are winner-take-all (WTA) networks~\cite{flor2014}.


\subsection {Learning Mechanisms}
\begin{figure}[h]
	\centering
	\includegraphics[width=1\columnwidth]{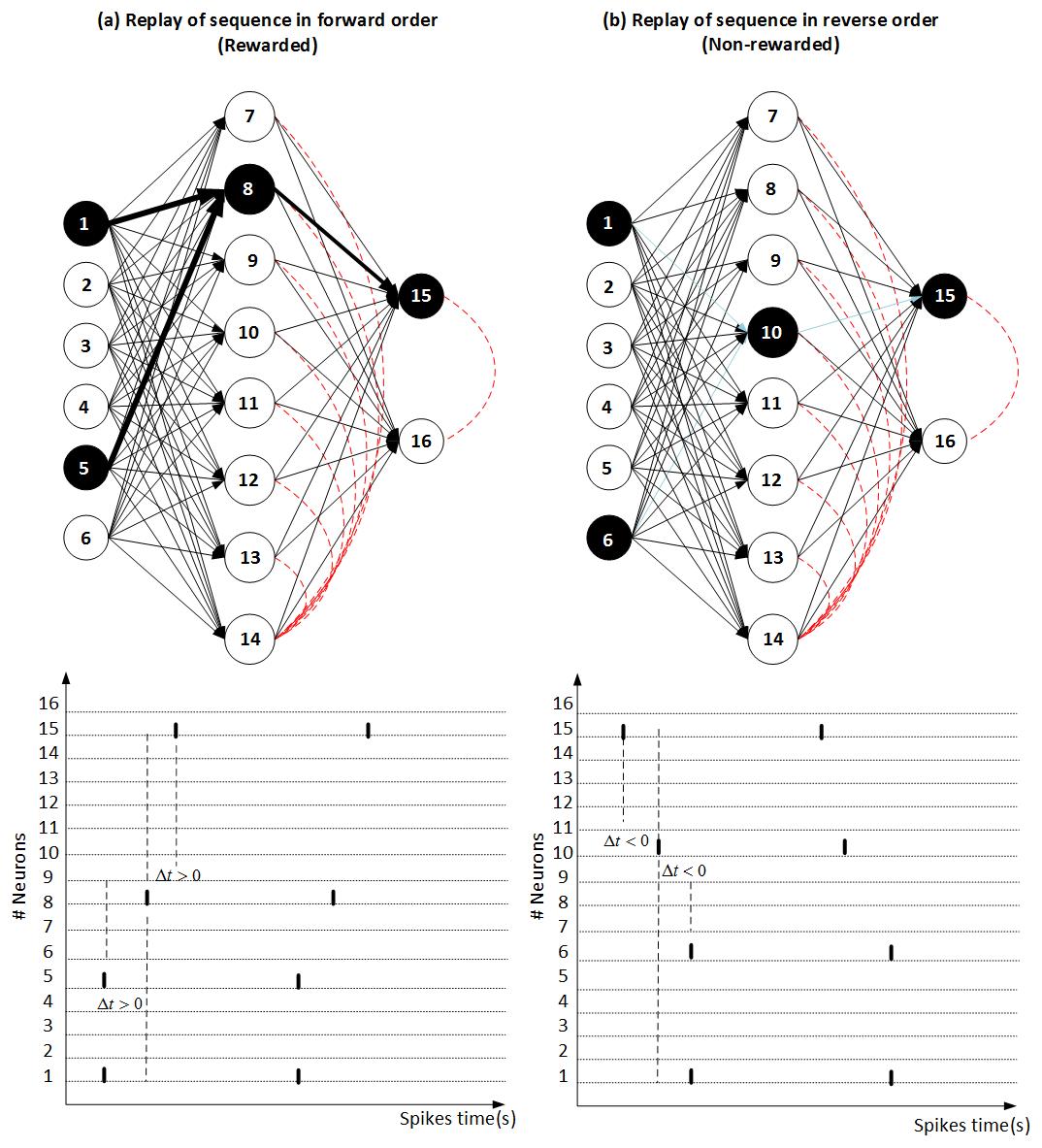}
	\caption{(a) Replaying the rewarded action sequence: triplet A1X, action dig; neurons in the first layer fire first then the second layer and finally third layer fire respectively. (b) Replaying the non-rewarded action sequence in reverse temporal order: triplet A1Y, action dig; Firstly The third layer fires, then the second layer and the first layer fire respectively.
 } 
	\label{replay_fr}
\end{figure}

In this network, Hebbian learning is used as the basic learning mechanism,  modeled through an STDP rule combined with an activity replay. The learning approach for the context-dependent task is reinforcement learning. In RL, the agent receives rewards by interacting with the environment and updates its ``value function'' and/or ``policy'' (modelled by the feedforward weights in the network in Fig.~\ref{SNN_model}) by going through a sequence of state-action pairs. In the model, there are two phases during training: in the behavior phase in each trial a sequence of states and actions is stored in memory. Then, in the replay phase this state-action sequence is replayed in a forward order if the reward was received and in reverse order if no reward was received~\cite{flor2014}.
Fig.~\ref{replay_fr} shows how the learning mechanism works for both forward and reverse order replay. In Fig.~\ref{replay_fr}(a) the triplet A1X causes ``digging''\footnote{``Dig'' in the experiment is an action that the rat takes if it decides that there is a reward hidden in the currently selected box. Otherwise, the rat walks to the different location, an action termed ``move'' in the experiments.} and the agent receives a reward. During replay of the rewarded state-action sequence, the input layer fires first, the second layer fires next and finally the output layer fires. This causes a positive timing difference for all related synapse and LTP\footnote{Long-term potentiation, i.e. increase of the synaptic weight.}. Digging in triplet A1Y causes a non-rewarded action which is shown in  Fig.~\ref{replay_fr}(b). During replay in reverse temporal order, firing starts from the third layer and the first layer fires at the end. In this case, all post-synaptic neurons fire before pre-synaptic neurons, leading to LTDs\footnote{Long-term depression, i.e. decrease of the synaptic weight.} in the synapses~\cite{flor2014}. 

\section{System Architecture}
\begin{figure*}[h]
	\centering
	\includegraphics[width=1.7\columnwidth]{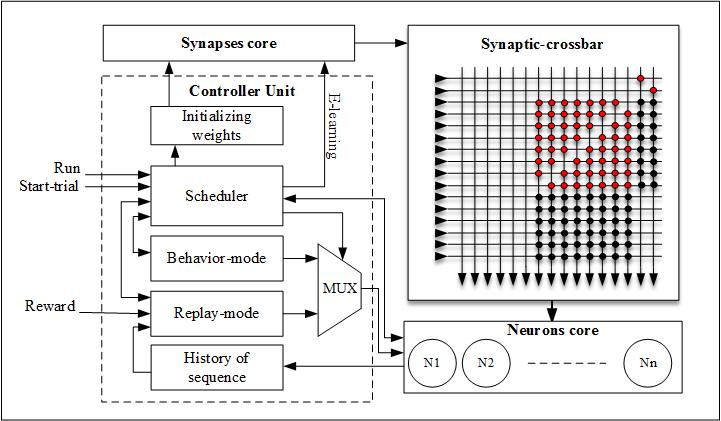}
	\caption{System level architecture of digital multiplier-less event-driven SNN for learning a context-dependent task. } 
	\label{SNN_BD}
\end{figure*}

The proposed architecture implements spiking neurons connected  by the plastic and static synapses through the synaptic crossbar. Neurons communicate with each other by sending spikes. We design the network architecture to be  event-driven and extremely parallel. There are multiply operations in the dynamic equation of neuron and synapse models behaviour.  A multiplier is the most expensive core in these modules in terms of area, latency, and power consumption. Thus, in the proposed hardware model, we aim to avoid using any multiplier.\\
Fig.~\ref{SNN_BD} shows the architecture of the proposed fully digital SNN. Similar to the other neuromorphic systems, spiking neurons and synapses are the most important parts of the proposed system. In this network, a synaptic crossbar connects all neurons. Besides neurons, synapses and synaptic crossbar, some peripheral cores for managing network behavior are crucial. Therefore the system consists of four different parts: a \texttt{neurons} core that stores all neurons from different layers in a row, a \texttt{synapses} core, a  \texttt{synaptic-crossbar} core that provide appropriate connections among all neurons, and finally a \texttt{controller-unit} for managing all sequences. We will present the details of each block in the following.

\subsection{Neurons Core}
Neuron is the main computing element in a SNN. All neurons in the neuron core are sorted in a row. The neuron model used in our hardware is based on a low-complexity LIF. The dynamic behavior of the neuron is similar to a first order circuit that only contains one storage element. The membrane potential $ V_{m} $ is modelled by electrical  potential on a capacitance C driven by the input current I while there is a constant leakage current through $G_{l}$ channel. In the LIF equation, $ V_{reset} $ is the resting potential for the membrane~\cite{Gerstner2002,flor2014}. Additionally, when the membrane voltage $V_{m}$ crosses the threshold voltage, $ V_{th} $, the neuron fires a spike at the output and, in the next time step, the membrane voltage is set to $ V_{reset} $~\cite{Gerstner2002, Andrew2013}. In the following we present an in-depth description of our implementation of the LIF neuron model.  


\subsubsection{Digital Neuron Implementation}
\begin{figure}[h]
	\centering
	\includegraphics[width=0.6\columnwidth]{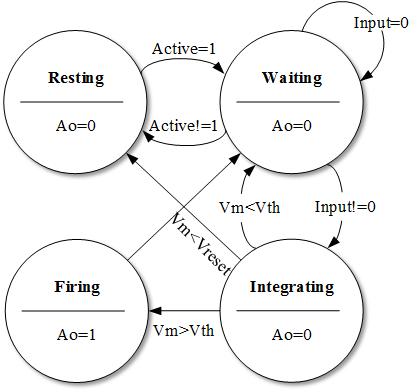}
	\caption{The state diagram of the LIF neuron. } 
	\label{fig:SM}
\end{figure}

\begin{figure}[h]
	\centering
	\includegraphics[width=1.0\columnwidth]{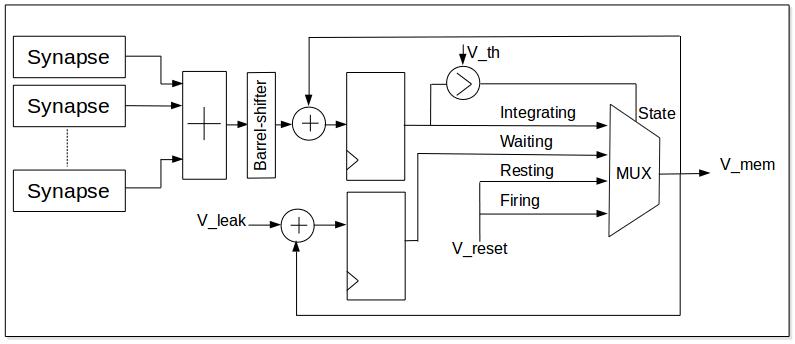}
	\caption{Circuit level block diagram for calculating the membrane voltage of neuron in each state. } 
	\label{fig:SM_BD}
\end{figure}

Operation of LIF neuron model can be simplified in four basic states. 1. Resting, 2. Waiting, 3. Integrating, 4. Firing. The LIF neuron model is summarized in Table~\ref{Tab:SM_BD}. In the following, the neuron operation in each state is described:
\begin{itemize}
\item \textbf{Resting-state}: In this state the neuron is not active and the membrane voltage is equal to resting voltage. According to the  Table~\ref{Tab:SM_BD}, the membrane voltage for an inactive neuron is equal to $V_{reset}$.
\item \textbf{Integrating-state}: At the arrival of input spikes, neuron's state changes to integrating. For the $j^{th}$ neuron in the $n^{th}$ time step, the membrane potential $V_{m}^{j}[n]$ is the sum of the membrane potential in the previous time-step $V_{m}^{j}[n-1]$ and the synaptic input. For each of the N synapses, the synaptic input is the spike input $A_{i}$ (input spike from the $j^{th}$ neuron) multiplied by the signed synaptic weight $W$. $W[i][j]$ is the synaptic weight between the  $i^{th}$ and $j^{th}$ neurons. In this model, it is possible to provide a spiking rate, expressed as input potential $V_{in}^{j}$ directly as input to neurons instead of individual spikes. Following each Integrating state, the LIF neuron state can change to either waiting or firing states based on the membrane voltage value. 
\item  \textbf{Waiting-state}: In this state the leakage value  $V_{l}$ is subtracted from the membrane potential. With a linear leak, this constant value is subtracted every  time-step regardless of the membrane potential value. In this state the LIF neuron is active and is waiting for the next input spike or input constant voltage.
\item \textbf{Firing-state}: Whenever the membrane voltage reaches the threshold, the value state changes to firing-state. In this state a spike is generated at the neuron's output, the  membrane voltage is set to the resting value, and the neuron goes to the waiting-state again.
\end{itemize}
As a neuron's output is a function of only its current state and not its input,  the proposed neuron model is based on a Moore Finite State Machine concept. 
The state diagram of the neuron which describes visually the operation of the circuit is depicted in Fig.~\ref{fig:SM}. Every circle represents a state in the upper part and the output of the neuron in the lower part. Every arrow represents a transition from one state to another. A transition happens once every clock cycle. Depending on the current Input, we may go to a different state each time. In the integrating-state, the transitions happen due to the current membrane voltage. The circuit level block diagram of membrane potential dynamical behavior of the neuron in each state is depicted in Fig.~\ref{fig:SM_BD}. \\

\begin{table}[h]
	\renewcommand{\arraystretch}{1.1}
	\centering
	\caption{The LIF neuron equations}
\label{Tab:SM_BD}
	\begin{tabular}{|l |l|}
	\hline 
	\multirow{2}{*}{\textbf{Resting}}
                & No-spike\\ 
                & $V_{m}^j[n]=V_{reset}$ \\\hline
    \multirow{2}{*}{\textbf{Integrating}}                
                & No-spike\\ 
                & $V_{m}^j[n]=V_{m}^j[n-1]+\sum_{i=1}^{all} A_{i}W[i][j]+V_{in}^j[n]$ \\\hline
    \multirow{2}{*}{\textbf{Waiting}}                
                & No-spike\\
                & $V_{m}^j[n]=V_{m}^j[n-1]-V_{l}$ \\\hline    
    \multirow{2}{*}{\textbf{Firing}} 
                & Spike \\
                &$V_{m}^j[n]=V_{reset}$ \\
    \hline 
    \end{tabular}
\end{table}

\subsection{Synapses Core}
The other most important unit of a spiking neural network is the synapse. In this network there is a couple of synapse types: inhibitory static synapses and excitatory plastic synapses. In the first group, the synapses have a negative effect on membrane voltage and provide lateral inhibition within neurons. All inhibitory synapses provide a strong inhibition and their value does not alter during network operation.  In contrast to inhibitory synapses, the excitatory synapses enhance the postsynaptic neuron's membrane voltage. The values of these synapses are modified during replay-mode due to the spike-timing-dependent plasticity (STDP) learning algorithm. In the next subsections, the learning algorithm is described in detail.
\subsubsection{Plastic Synapse Dynamics}
The proposed computational model uses STDP based rule for the weight update~\cite{flor2014}. The STDP algorithm adapts the weight of a synapse according to timing difference between the pre- and post-synaptic spikes arrival times. If a pre-synaptic spike arrives several milliseconds before the post-synaptic spike, the weight of the synapse will be raised. This gives a positive time difference $\Delta t>0$, which leads to a synaptic LTP. On the other hand, the weight will be decreased if the post-synaptic spike arrives before than pre-synaptic spike. This gives a negative time difference $\Delta t<0$, which leads to synaptic LTD~\cite{Bi1998}. The amount of synaptic modification is determined by the following equations:

\begin{equation}
\tau_{w}\frac{dW}{dt} =
\begin{cases}
A^{+}(W_{Max}-W)exp(\text{$\Delta t $}/\text{$\tau$}_{+}) & \text{ $\Delta t>0, $} \\

A^{-}(W_{Min}-W)exp(\text{$\Delta t $}/\text{$\tau$}_{-}) & \text{ $\Delta t<0, $ }
\end{cases}       
\end{equation}
Where $W$ is the synaptic weight and $W_{Max}$ and $W_{Min}$ determine the dynamic range of its variation, $\frac{dW}{dt}$ is the modification rate of the synapse, $\Delta t$ is the timing difference between arrival time of the pre- and post-synaptic spikes, $A^{+}$ and $A^{-}$ are amplitudes of synaptic modification and the time constants $ \tau_{+}, \tau_{-} $ and $ \tau_{w} $ control the weight adaptation, exponentially decaying influence of LTP and LTD, respectively. Such effects occur within a small time window. Fig.~\ref{fig:STDP}(a) depicts the modification value of the synapse over timing difference for three different initial synapse values. In the background computational model, the amplitude for potentiating is three times bigger than the amplitude for depression~\cite{flor2014}.

\subsubsection{Digital Synapse Implementation}
\label{sec:Synapse}
\begin{figure}[h]
	\centering
	\includegraphics[width=1.0\columnwidth]{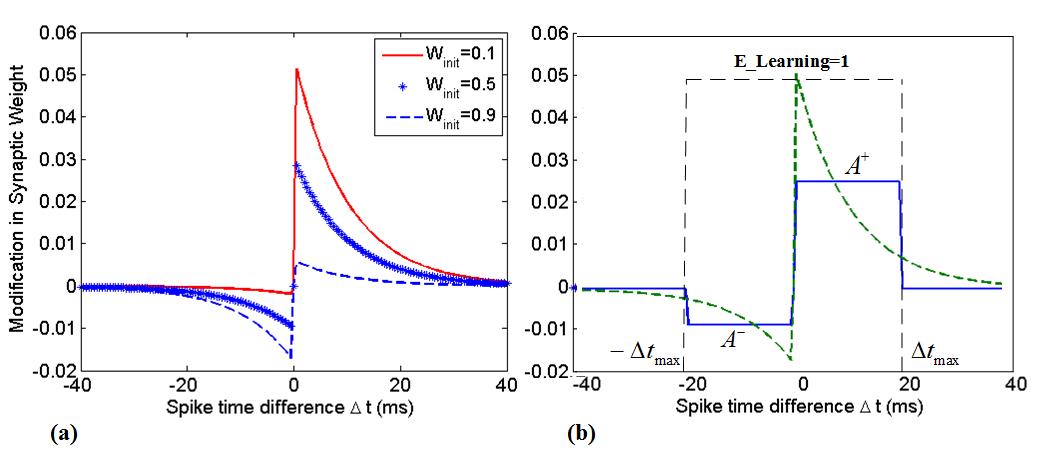}
	\caption{(a) Synaptic modification of original rule over spike time difference for three initial weights. (b) Original (green dashed line) and proposed simplified (blue line) STDP modification function over timing differences between arrival of pre- and post-synaptic spikes. } 
	\label{fig:STDP}
\end{figure}

For implementing synapses on silicon the adaptation rule should be simplified. Otherwise, implementing the original learning rule with exponential functions and multipliers causes utilization of a large silicon area~\cite{Run2016, Cassidy2011}. In synapse model of the proposed system, synaptic modification value depends on the previous strength of synaptic weight. Effects of current synapse weight on the next synapse weights is very important in order to achieve the correct performance. For maximum simplification and to keep this dependency, we propose the following simple learning rule:  
\begin{equation}
\Delta W =
\begin{cases}
A^{+}(W_{Max}-W) & if \text{ $\Delta t>0 $ and $E_{learning}=1$,} \\ 

A^{-}(W_{Max}-W) & if \text{ $\Delta t<0 $ and $E_{Learning}=1$,}
\end{cases}       
\end{equation}
where the  $\Delta W$ is the synaptic modification value. Whenever the scheduler sends a signal to enable the learning phase ($E_{learning}=1$) the synapse is eligible to adapt.
Fig.~\ref{fig:STDP}(b) shows the comparison between the proposed simple STDP based synapse and the original model ($W_{init}$=0.5).\\ 
\begin{algorithm}
    \caption{Simplified Plastic Synapse Algorithm.}\label{alg:synapse}
    \hspace*{\algorithmicindent} \textbf{Input} pre and post-synaptic spikes. \\
    \hspace*{\algorithmicindent} \textbf{    }     $E_{learning}$, $W_{Previews}$ \\
    \hspace*{\algorithmicindent} \textbf{Output} $W_{new}$ 
    \begin{algorithmic}[1]
    \If {(Spike pre = True)}  
        \State Store $T_{pre} $ 
    \EndIf
    \If {(Spike post = True)}  
        \State Store $T_{post} $ 
    \EndIf
\While{$E_{learning} = True$}  \Comment{Eligible to adapt}
        \State $\Delta t \leftarrow T_{pre}-T_{post}$  
    \If{$\Delta t > 0$}
        \State $W_{new} \leftarrow W_{Previews}+ (W_{Previews} << A^{+})$
        
        \ElsIf{$\Delta t < 0$}
        \State $W_{new} \leftarrow W_{Previews}- (W_{Previews} << A^{-})$
    \Else
        \State $W_{new} \leftarrow  W_{Previews}$
       
    \EndIf
        
    \EndWhile  \label{roy's loop}
    
    
    \end{algorithmic}
    \end{algorithm}
Algorithm~\ref{alg:synapse} describes the proposed simplified hardware model for this synapse. As long as the synapse is eligible for learning, i.e. {$E_{learning}= True$}, synaptic modification is calculated based on the pre- and post-synaptic spikes timing difference. 
A necessary procedure before entering the algorithm is initializing synapses weights. Initial weights for excitatory plastic synapses in this model have positive values around $(W_{Max}-W_{Min})/2$. To generate these numbers, several configurable linear feedback shift register blocks (LFSR) are employed. On the circuit level, an LFSR consists of a few flip-flops in a row, wired as a shift-register with feedbacks through XOR gates~\cite{Navabi2011}. Although there are a lot of models for LFSRs, to study this network with different initial synaptic weights, configurable LFSR was selected. A configurable LFSR has this ability to start random numbers from an arbitrary number as a seed and has a random polynomial equation as feedback~\cite{Navabi2011}. 

\subsection{Synaptic Crossbar}
In this network, as shown in Fig.~\ref{SNN_BD}, all neurons are sorted in a row and are connected through the synaptic crossbar. In the synaptic crossbar in Fig.~\ref{SNN_BD}, the excitatory and inhibitory synapses are illustrated using black and red circles, respectively. 
In this system, WTA networks are created using positive and negative synaptic weights (excitatory and inhibitory synapses). Neurons in each layer amplify their local activity by being excitatory connected to neurons in the previous layer. In the WTA networks, the winners inhibit their neighboring neurons by being all connected via inhibitory synapses. 
This type of WTA computation has been
proposed in hierarchical models of vision and in models
of selective attention and recognition~\cite{Rapha2017, Ries1999,Sandamirskaya2014}. \\
\begin{table}[h]
	\renewcommand{\arraystretch}{1.2}
	\centering
	\caption{Details of selected parameters}
	\label{parameters}
	\begin{tabular}{ l l l  }
		\hline
		
		\textbf{LIF neuron}     & &     \\ \hline
		Threshold membrane potential & $ V_{th} $      &  -50 mV   \\ 
		Reset membrane potential     & $ V_{reset} $   &  -70 mV   \\ 
		Neuron leakage voltage       & $ V_{leakage} $ &  $1.2\times 10^{-7}$(V)   \\  \hline
		\textbf{STDP rule} & &    \\  \hline
		Spike amplitude when $ \Delta t>0 $ & $ A{+} $       &  $+2^{-10} $   \\ 
		Spike amplitude when $ \Delta t<0 $ & $ A{-} $       & $-2^{-11} $   \\ 
		Synaptic weight maximum activation  &  $ W{max} $    & 1   \\ 
		Synaptic weight minimum activation & 	$ W{min} $     &  0   \\ \hline
		\textbf{Design choice}   &      &    \\  \hline
		Number representation    & $ Nb  $      &  32 fixed-point   \\ 
		$\# $ Fractional bits    & $ Nb_{f} $   &  31   \\ 
		$\# $ Integer bits       & $ Nb_{I} $     &  1   \\ 
		$\# $ Input neurons      & $N_{Input}$    &  6   \\ 
		$\# $ Hidden neurons     & $N_{Hidden}$    &  8   \\ 
		$\# $ Output neurons     & $N_{Output}$    &  2   \\
		Maximum time interval for a trial          & $ T_{trial} $ &   30000 $T_{clk}$   \\
		Maximum time interval for replay          & $ T_{replay} $ &   130 $T_{clk}$   \\
		Input voltage for input neurons          & $ V_{input} $ &   1.28 mV   \\
		Input voltage for hidden neurons          & $ V_{hidden} $ &  1.48 mV   \\
		Input voltage for output neurons          & $ V_{output} $ &   1.64 mV   \\
	 \hline
	\end{tabular}
\end{table}
\subsection{Controller Unit}
The controller unit is responsible for controlling behavior of the system, sequences of states, data storage, and preparing initial weights for synapses. The controller unit consists of five main sub-blocks: 
\begin{itemize}
\item \textbf{Scheduler}: responsible for controlling the network sequences. This block receives controlling signals, spikes of neurons, and feedback from other controlling blocks. Based on the input signals, the scheduler decides the next system operation and sends controlling signals to neurons, synapses, and other controlling blocks.
\item \textbf{Behaviour-mode}: for controlling the SNN inputs during the behavioral phase. This block gets inputs from the scheduler and is always aware of the neurons' spikes. This block is responsible for providing suitable  input to the neurons in the behavior phase, sending the neurons activities to the history block, and sending back the controlling signal to the scheduler.
\item \textbf{Replay-mode}: for managing network parameters during the replay and learning phases. This block is controlled by the scheduler and provides suitable inputs to neurons during the replay phase and also sends back the controlling signal to the scheduler.
\item \textbf{History-of-sequences}: for saving the sequence of two latest neurons activities in the behavioral phase. This control block receives inputs from the Behaviour-mode block, saves them for a set time and provides the required information for the Replay-mode block.
\item \textbf{Initializing-synapses}:  generates all initial inhibitory and excitatory values for synaptic weights using linear feedback shift register (LFSR) blocks. This block gets a controlling signal from the scheduler and sends initial weights to the synaptic crossbar block.
\end{itemize}
\section{Results and Discussion}
\subsection{Modeling the Biological Experiment in Hardware}
The spiking neural network model for learning context-dependent task includes 16 neurons (6 input, 8 hidden and 2 output neurons), 64 plastic excitatory synapses ($6\times8 $ synapses between input and hidden layers and $8\times2 $ synapses between hidden and output layers), 58 inhibitory non-plastic synapses (56 for hidden layer and 2 for output layer) and, finally, the controller parts. 
At the beginning of the experiment, all the neurons are in the resting state.  The initializing-synapses core provides initial random weights for plastic synapses. This core also constructs WTA networks by assigning strong inhibition value to the appropriate synaptic connections. Network operation repeats more than 100 trials and each time network starts from a randomly selected input stimulus. Each trial consists of two different modes: behavioral and replay. The controller unit receives a controlling signal \texttt{run} for starting whole experiment and also \texttt{start-trial} for starting each trial. The scheduler also receives spikes from the output layer of the neuronal network. At the beginning of a trial, the scheduler enables the behavior-mode core. According to spikes from the output layer, after getting events from the ``dig'' output neuron, the scheduler switches to the replay mode. The scheduler also assigns a selective signal to the multiplexer which provides input voltages for neurons in each mode. Operation of the network in both modes is described in the following: 
\subsubsection{Behavior-mode}
During the behavior-mode, the network starts from a randomly selected triplet, for example, starts from context A, Position 1 where item X is located (A1X). As each trail occurs in one of the two contexts, each triplet has an accompanying complimentary combination of the context, position, and item. For example, A2Y is a complimentary of A1X. Based on the value of the synaptic weights, the behavior phase can include a few movements between the two triplets (if ``move'' output neuron is active) and ends up by digging on one of them (if ``dig'' output neuron is active). 
After each movement or each dig, the states of all neurons are sampled in the history-of-sequences. The goal is that by the end of the behavior phase, the history of a maximum two latest actions sequences (stimulus-response pairs) is stored. 
\subsubsection{Replay-mode}
In the replay phase, the scheduler sends controlling signal \texttt{E-learning} to the plastic synapses and makes them eligible to learning. Furthermore, according to the obtained reward and the sampled action sequences, the Replay-mode unit provides suitable inputs for all neurons. The sequences replay within a specific time window (Table~\ref{parameters}). The rewarded action sequences  replay in a forward temporal order. In order to encourage them to be chosen in the future, combined with the STDP learning rule this replay enhances the related synapses. Non-rewarded action sequences replay in the backward temporal order. Combined with the STDP learning rule such replay discourages the future use of such action sequences. 
\subsection{Implementation Results}
\label{sec:results}
In this section we present simulation and implementation results for the proposed network.
All neurons, STDP-based synapses and other peripheral parts are described using the standard top-bottom digital ASIC design flow with separate fully-synthesizable synchronous modules and therefore can be implemented with state-of-the-art manufacturing techniques. As multiplier is a high-cost core in terms of area utilization and power consumption, the proposed architecture doesn't contain any multiplier. The Design choices we made are presented in Table~\ref{parameters}. In order to get a fair balance between accuracy and cost, the proposed spiking neural network model performs data transformation using 32-bit fixed-point numbers with 31 fractional bits. As this research doesn't aim to consider reinforcement learning for large scale spiking neural network, there are not many neurons and synapses in this network, therefore we made point to point connections. In order to minimize silicon area utilization, neuron and synapse blocks are designed as simple as possible. 
As a proof of concept we implement the network on Kintex-7 XC7kt160t FPGA which is hosted in opal kelly XEM7360 board.  Table~\ref{comp:neuron_result} and Table~\ref{comp:synaps_result} present the FPGA device utilization of the proposed neuron and synapse implementation in comparison with the previously published work. Kintex-7 XC7kt160t chip totally contains 101400 slice LUTs and 202800 slice Flip Flops. Each neuron uses less than 0.01 $ \% $ of available slice FF and 0.29 $ \% $  of whole slice LUTs. Each synapse uses 0.02 $ \% $  and 0.17 $ \% $  of slice FF and slice LUTs respectably. Table~\ref{comp:result} reports the FPGA device utilization of the proposed SNN and a collection of previous studies on implementing different SNN models, in which spiking neural networks are implemented on similar hardware platforms. Please note that these neurons, synapses, and networks are not applied in the same task. Furthermore, different FPGA devices and synthesizer versions have been used for implementations. Therefore, the device utilization results presented in these tables must be considered with caution. As Table~\ref{comp:result} shows, the proposed spiking neural network totally uses 4.39 percent of available Slice FFs, 18.80 percent of whole available slice LUTs. 
\begin{table*}[h]
	\renewcommand{\arraystretch}{1.1}
	\centering
	\caption{Device utilization plus performance comparison for the implemented neuron module. Abbreviations Used in this table correspond to the circuit used for each neuron model. Abbreviations for this table are: Izhikevich (Izh), Adaptive-Exponential (Adex), Morris-Lecar (MOL), FitzHigh-Nagumo(FINA), Hindmarch-Rose (HR).}
	\label{comp:neuron_result}
	\begin{tabular}{ l l l l l l l l l l }
		\hline
		Resource       &\cite{Sol2012}&\cite{Gom2014}&\cite{Heidar2016}& \cite{Hay2015}      & \cite{Nouri2015} & \cite{Hay2016}         & \cite{Karimi2018}     & \cite{Edris2019}  & This work   \\ \hline
		Neuron Model  &Izh         & Adex &Adex & MOL&FINA  &HR& Wilson   & LIF    & LIF   \\
	    Slice Registers& 0         &  0   & 829 & 0           & 0               & 0           & 365      & 46      & 0  \\
	    Flip Flops     & 493       &  388 & 0   &618          &  526            & 431         & -        & 0       & 2  \\ 
		Slice LUTs     & 617       &1279  &1221 & 3616        & 1085            & 659         & 611      & 56      & 292 \\ 
		Max speed(Mhz) & 241.9     & 190  &134.3& 135         & -               &  81.2       &  98      &  412.37 & 149 \\
		Device         &Virtex-II  &Virtex-II&Spartan6&Virtex-II&Virtex-II&Virtex-II&Virtex-6&Virtex-6&Kintex7  \\
		\hline
	\end{tabular}
\end{table*}

\begin{figure*}[t!]
    \centering
    \begin{subfigure}{0.3\textwidth}
        \centering
        \includegraphics[height = 0.75\textwidth]{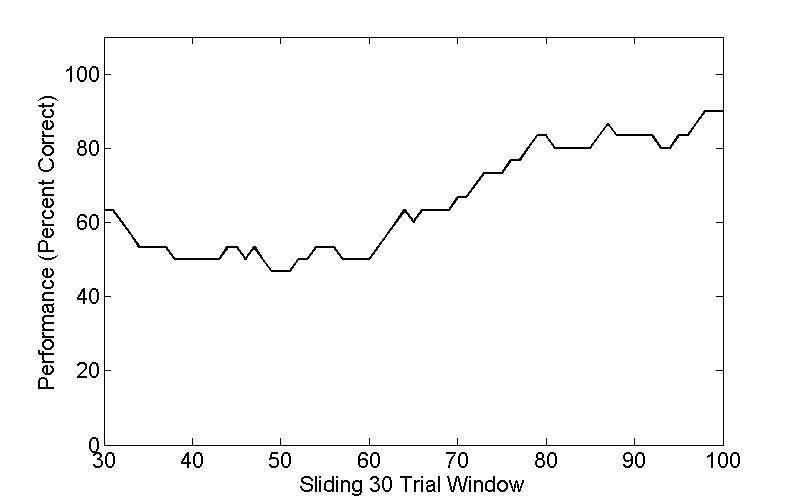}
        \caption{}
    \end{subfigure}%
    \qquad
    \begin{subfigure}{0.3\textwidth}
        \centering
        \includegraphics[height = 0.75\textwidth]{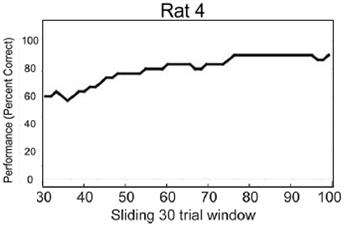}
        \caption{}
    \end{subfigure}
        \qquad
    \begin{subfigure}{0.3\textwidth}
        \centering
        \includegraphics[height=0.75\textwidth]{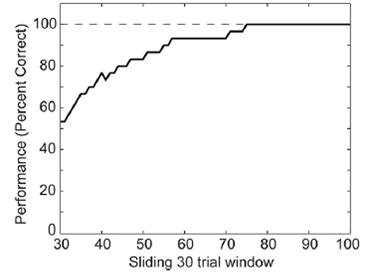}
        \caption{}
    \end{subfigure}
    
    \caption{Behavioral performance during successful learning of the context-dependent task: (a)~The proposed digital  event-driven hardware model. (b)~The animal experiment~\cite{Komo2009}. (c)~The computational model~\cite{flor2014}.}
    \label{performanc}
\end{figure*}

\begin{table*}[h]
	\renewcommand{\arraystretch}{1.1}
	\centering
	\caption{Device utilization plus performance comparison for the implemented synapse module. Abbreviations Used in this table correspond to the circuit used for each plasticity rule. These abbreviations include: Pair-based STDP (PSTDP), Minimal TSTDP Hippocampal (MTH), Minimal TSTDP Visual Cotrex (MTVC), Calcium-Based (CAB),  Serial PSTDP (Serial), Cell-based PSTDP (Cell-based).}
	\label{comp:synaps_result}
	\begin{tabular}{l l l l l l l l l l l l}
		\hline
		Resource       &\cite{Cassidy2011}&\cite{Jok2017}& \cite{Nou2017}& \cite{Nou2017}& \cite{Nou2017}& \cite{Bilel2009} &\cite{Bilel2009}     &\cite{Run2016}&\cite{Mostafa2019}&\cite{Mostafa2019}& This work   \\ \hline
		Learning Rule  &PSTDP   & CAB  &PSTDP& MTH  &MTVC  &Serial&Cell-based&PSTDP&PSTDP&PSTDP& PSTDP   \\
	    Slice Registers& -      & -    & 46  & 54   &  47  & -    & -        & -   & 12  & 642 &-\\
	    Flip Flops     & 39     &292   & -   & -    & -    & -    & -        & 398 & 16  & 671 &38\\ 
		Slice LUTs     & 18     &309   &36   & 41   & 26   & 47   & 339      & 1430& 8   & 859 &172\\ 
		Max speed(Mhz) & -      & 332  & 138 & 192  & 192  & -    & -        &  200& 816 &362  & 151\\
		Device         &Spartan3&Spartan6 &Spartan6&Spartan6&Spartan6&Spartan3&Spartan3&Virtex6&  Spartan6&Spartan6&Kintex7  \\
		\hline
	\end{tabular}
\end{table*}
\begin{table*}[h]
	\renewcommand{\arraystretch}{1.1}
	\centering
	\caption{Device utilization plus maximum frequency comparison for the implemented SNN.}
	\label{comp:result}
	\begin{tabular}{l l l l l l l l l}
		\hline
		Resource         &\cite{Edris2019} &\cite{Bonabi2014}           &\cite{Neil2014}&\cite{Edris2015}         &\cite{Darwin2017}          &\cite{Lammie2018}     & This work   \\ \hline
	    Slice Registers & 1023   &50228 &-   &1032  &1676 &33  &-  \\ 
	    Flip flops      & -      &-     &-   &-     &-    &-   &8906  \\ 
		Slice LUTs      & 11339  &86032 &-   &6264  &6214 &19  &19059  \\ 
		DSPs            & 0      & 1112 & -  &-     &32   &0   &0 \\
		Max speed(Mhz)  &189     &63.389& 75 &21.207&25   &717 & 148.4\\
		Device &Virtex6 &Virtex7 &Spartan6& Zynq & Spartan6 & Terasic&Kintex7  \\
		\hline
	\end{tabular}
\end{table*}
By implementing the system on an FPGA running at 100\,MHz ($T_{clk}$), we have a quite fast network. 
We simulate our event driven network with randomly initialized synaptic weights using the Xilinx Vivado Design suite. All the simulation results data were stored in different files and analyzed via MATLAB~\cite{senselab2014}. At the initial trials, the network could not distinguish the correct response. After several trials the network learned the task and ``digged'' When sensing either item X in context A or item Y in context B. 
Fig.~\ref{performanc}(a) depicts performance of the network in a number of trials. Performance is calculated by measuring the percentage of the mean number of correct responses over a sliding window of 30 trials. We proposed an event-driven hardware model which is able to learn the same task as in biological experiments and previous computational studies within about 100 trails and reached the performance within 80 $ \% $ to 90 $ \% $ (Fig.~\ref{performanc}(a)). The performance converges after 100 trial and doesn't change if training continuously. As shown in Fig.~\ref{performanc}(b) in the experimental study the context-dependant task was learned in about 100 trials and rats reached about 90$ \% $ correct behavioral response~\cite{Komo2009}. The computational model was able to learn the task in about 100 trials in average 80$ \% $ to 90$ \% $ correct detection rate~\cite{flor2014} (Fig.~\ref{performanc}(c)). Thus, the proposed digital event-driven network has the performance more similar to animal behavior in the experiment.

\begin{figure*}[h!]
	\centering
	\includegraphics[width=1.7\columnwidth]{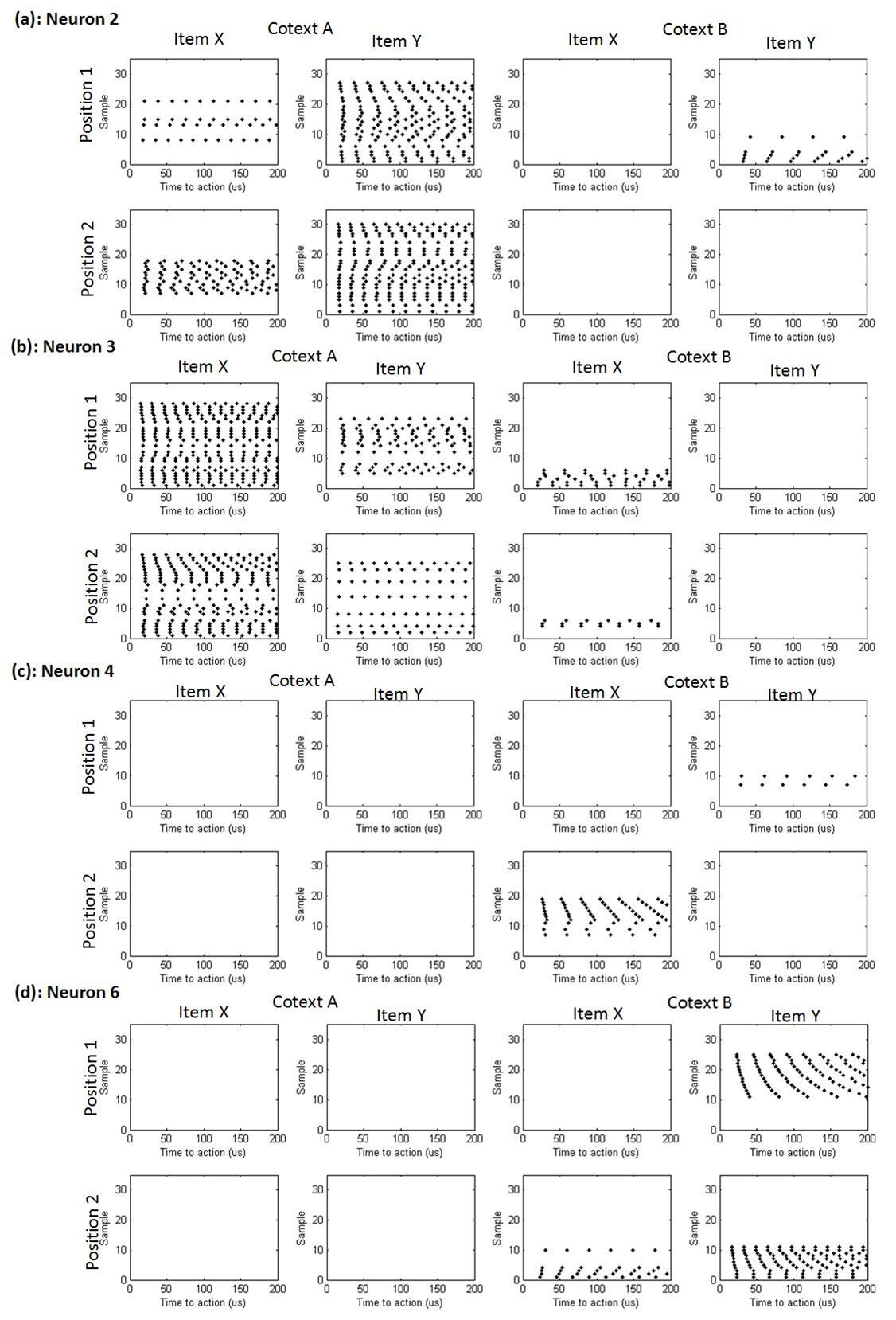}
	\caption{Evolution of spike events for four functional neurons out of eight hippocampus neurons. Firing patterns are shown for (a) neuron $ \#$2, (b) neurons $ \#$3, (c) neurons $ \#$4 and (d) neurons $ \#$6. The sample index- these are the rows in the spike raster plots- appear in the same temporal order as they have been recorded per panel. 
	} 
	\label{RasterPlot} 
\end{figure*}

\begin{figure}[h]
	\centering
	\includegraphics[width=1.0\columnwidth]{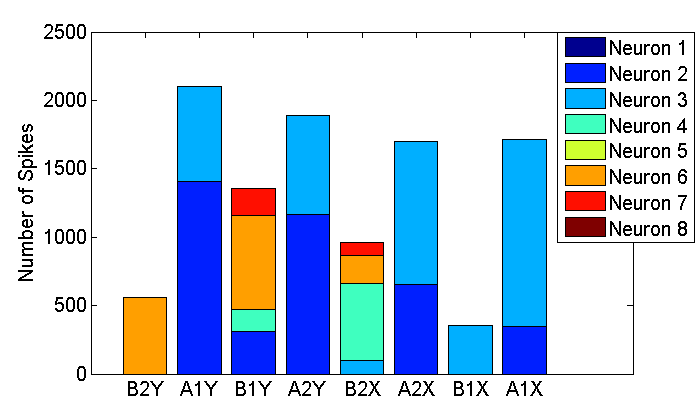}
	\caption{Number of total neurons spikes over each triplet. Neurons 2
fires selectively for item Y in context A, regardless of the place the
item Y appears. Neuron 3 fires selectively for item X in
context A. Neuron 6 fires selectively for item Y in context B. Neuron 4 fires selectively for item X in context B and place 1. } 
	\label{graph_bar}
\end{figure}

\begin{figure}[h]
	\centering
	\includegraphics[width=1.0\columnwidth]{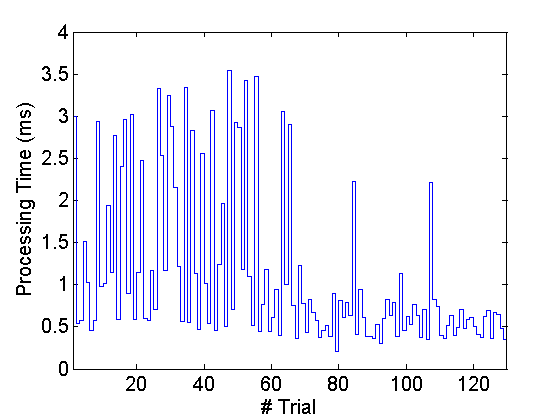}
	\caption{Network required processing time for each trial.} 
	\label{processing_time}
\end{figure}

We evaluated the spike patterns during learning the context-dependant task in a single exemplary run. In this model, each time four out of eight neurons are included in the functional network. In this specific run, as shown in Fig.~\ref{RasterPlot}, neurons~2, 3, 4 and 6 are mostly involved in the functional network. Fig.~\ref{RasterPlot}(a) depicts spiking events of neuron~2 for eight different triplets. This neuron starts firing mostly whenever network senses Item~Y in context~A regardless of which position Item~Y is located (Triplets A1Y and A2Y). As Fig.~\ref{RasterPlot}(b) shows, the neuron~3 mostly fires when network model sense item~Y in context~A (triplets A1X and A2X). Spikes of neurons~4 and 6  are depicted in Fig.~\ref{RasterPlot}(c) and (d). Neuron~4 fires whenever network sense Item X and neurons 6 fires whenever network senses Items  X and Y, both in context B (Triplets B2X, B1Y, and B2Y). The raster plot diagrams also confirm that in this run performance is nearly 90$\%$  and the hardware model finally could not find the reward in triplet B1X. Total number of spikes over each input triplet for each neuron are illustrated in Fig.~\ref{graph_bar}. Toward the end of the simulation, the neuron~2
fires selectively for item Y in context A, regardless of the place the
item Y appears  (Fig.~\ref{graph_bar}). Neuron~3 fires selectively for item X in
context A. Neuron~6 fires selectively for item Y in context B. Neuron~4 fires selectively for item X in context B and place 1. Similar firing patterns were observed both experimentally and in the computational model.
Fig.~\ref{processing_time} reports required processing time for this network. As mentioned, for each trial a triplet is specified as a starting point, and each trial ends up in a ``dig'' action. Processing time means the whole required time for each starting point to end up in a dig. In the proposed neuron model the integration time for neurons depends on the synaptic weights' values. As after learning the synaptic weights converge to margins, the processing time for the latest trials is less than for the initial trials. \\
		
\subsection{Verification by Robot Experiment}
\label{sec:robot}
\begin{figure*}[h]
	\centering
	\includegraphics[width=1.6\columnwidth]{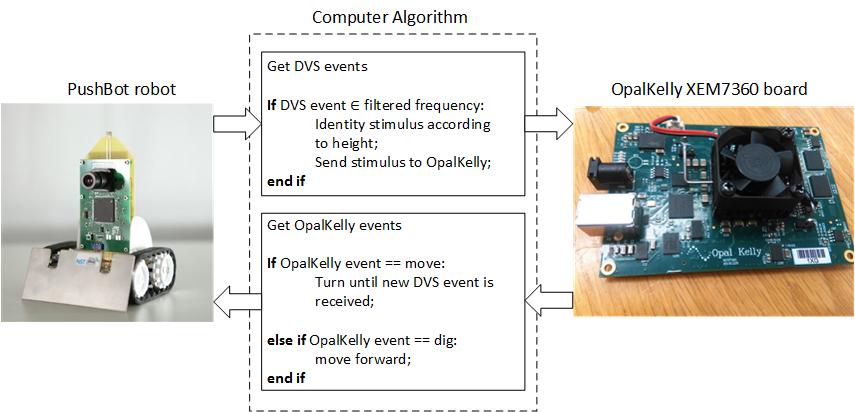}
	\caption{Overview of the digital event driven SNN for Hippocampus implemented on Opalkelly XEM7360 board and connecting to Pushbot robot setup} 
	\label{Experiment_Robot}
\end{figure*}

\begin{figure}[h]
	\centering
	\includegraphics[width=1\columnwidth]{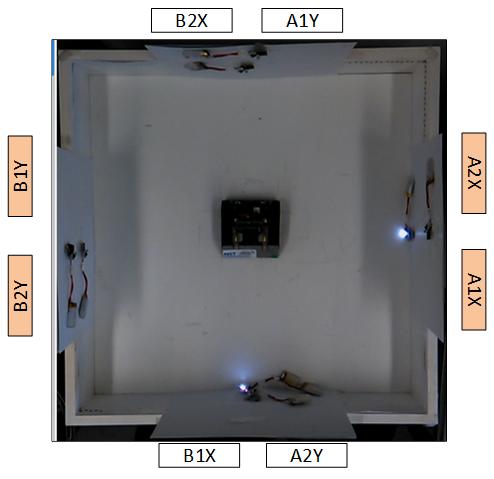}
	\caption{Hardware setup for repeating context dependant task experiment via digital neuromorphic chip. Eight Blinking LEDs are stuck in different heights and describe eight different triplets. Triplets with white sing block are non-rewarded (B2X, B1X, A1Y, A2Y) and Triplets with orange sing blocks are rewarded.  } 
	\label{Invironment_Robot}
\end{figure}

\begin{figure}[h]
	\centering
	\includegraphics[width=1.00\columnwidth]{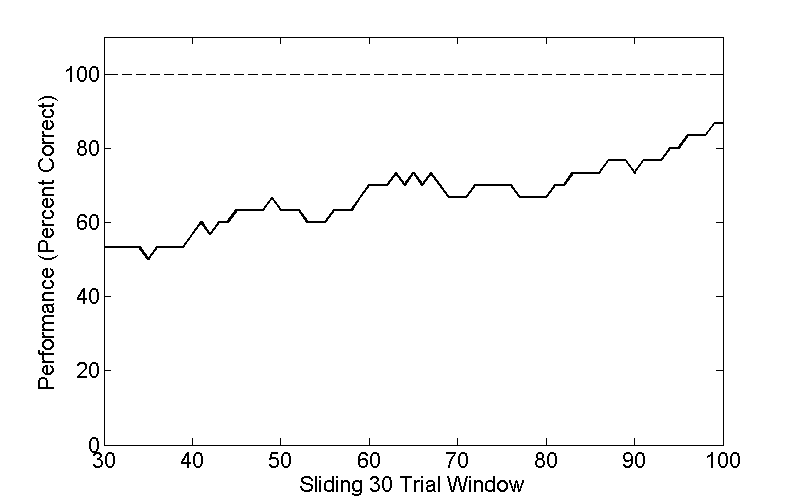}
	\caption{Behavioral performance of robot during learning the context-dependant task. } 
	\label{performanc_robot}
\end{figure}

In order to verify the proposed system, we interfaced the digital system on FPGA to a neuromorphic dynamic vision sensor (DVS) mounted on a robotic vehicle and developed an autonomous robotic agent that is able to learn a context-dependent task.
The experimental setup used in this work consists of the Pushbot robotic vehicle with an embedded DVS camera (eDVS~\cite{Conradt2009}) and the Opallkelly XEM7360 board. Furthermore, a computer is used to direct the flow of events between the robot and the SNN on FPGA.
Fig.~\ref{Experiment_Robot} shows the components of our hardware setup and the information flow between different hardware components.
The Pushbot communicates with the computer via a wireless interface for receiving motor commands and for sending events produced by the DVS~\cite{Moritz2017}.
The computer runs a simple program that manages the stream of events between the system on FPGA and the robot.
As shown in Fig.~\ref{Invironment_Robot}, we define different stimuli using LEDs placed at different heights (according to Table~\ref{Input_Robot}) and in different positions. Based on the detected LED, a reward is produced for the network internally. In each trial of the experiment, we put the robot in front of one of the LEDs as the starting point and detect the height of the LED based on the output of the DVS camera on the robot, which leads to an identification of the triplet. In the behavioral phase,  based on the network input and the current synaptic weights of the spiking neural network, the network decides to move or dig. In this verification set up, for moving the robot turns towards the other LEDs and for digging the robot drives closer to the LED. As shown in Fig.~\ref{performanc_robot}, similar to the simulation results, in robotic experiment performance starts from around 50\% in the first trials and gradually goes up to around 85\%. The supplementary movie shows robot behavior in first trials and also after complete training \footnote{Shared video on youtube: https://youtu.be/W8u7UEWyLAs}.\\

\begin{table}[h]
	\renewcommand{\arraystretch}{1.3}
	\centering
	\caption{Details of the experiment: Network inputs, LEDs' height, and expected actions after learning for each triplet.}
	\label{Input_Robot}
	\begin{tabular}{l c c c c c c c c }
		\hline
		
		  Triplet  &A1X&B1Y&A2X&B2Y&A1Y&B1X&A2Y&B2X     \\ \hline
		
		\multirow{6}{*}{Input}          & 1 & 0 & 0 & 0 & 1 & 0 & 0 & 0  \\ 
	         & 0 & 1 & 0 & 0 & 0 & 1 & 0 & 0  \\ 
	          & 0 & 0 & 1 & 0 & 0 & 0 & 1 & 0  \\ 
	          & 0 & 0 & 0 & 1 & 0 & 0 & 0 & 1  \\ 
	          & 1 & 0 & 1 & 0 & 0 & 1 & 0 & 1  \\ 
             & 0 & 1 & 0 & 1 & 1 & 0 & 1 & 0  \\ \hline
		Height\\(cm)    & 0.5 & 15 & 13 & 8  & 11 &  6.5 &2.5  & 19  \\\hline
		Dig         & 1 & 1 & 1 & 1 & 0 & 0 & 0 & 0  \\ 
		Move        & 0 & 0 & 0 & 0 & 1 & 1 & 1 & 1  \\ \hline

	\end{tabular}
	
\end{table}
\subsection{Discussion}
Most practical studies in the field of neuromorphic technology similar to neural network accelerator, have focused on pattern classification applications~\cite{Charlotte2018,Ning2015}. So far there has been little success on autonomous systems which learn tasks by interacting with the environment. A strong relationship between reinforcement learning and autonomous systems has been reported in the literature~\cite{Sutton2018}. In this study, we present a digital multiplier-less event-driven spiking neural network architecture for modeling reinforcement learning which is able to learn a context-dependent task. 
The performance of the proposed event-driven hardware in this task matches both the behavioral performance of the animal experiment and computational model~\cite{Komo2009, flor2014}. The hardware model, computational model, and also the robotic agent learned the task within 100 trials (see Fig.~\ref{performanc}). 
Although the structure of the model is relatively abstracted, it represents the interaction of the hippocampus region CA1 with sensory input and motor output. Based on the way that different parts of the hippocampus connect, the sensory layer could correspond to neurons in  region CA1 that receive a strong sensory input via the entorhinal cortex~\cite{Eich2007, Desh2009}. The motor output could correspond to neurons in region CA1 that have strong bidirectional connectivity with entorhinal neurons receiving representations of spatial actions from mPFC or parietal cortex~\cite{Hyman2010, Nitz2012, flor2014}. By FPGA implementation of the digital neural circuit with reinforcement capability and making a closed-loop connection between the FPGA chip and the robotic agent, we artificially recreate the animal experiment.
The model has a few parameters that are critical to the learning and convergence in the context-dependent task. In the replay phase, the input voltage value of neurons in each layer are very important and affect pre- and post-synaptic spikes' arrival timing differences. In order to prevent the network from overflowing, using enough bits to present numbers is an important factor.  In this network, we present all numbers by 32 bits.  Another important factor in this network is the number of random bits in initial synaptic weights which leads the WTA networks to operate more precisely. The value of LTP and LTD in plastic synapses are very important as well,  they need to prevent any saturation. There are two WTA networks -- in the hidden and output layers. To allow the hard WTA network to specify the winner more precisely, we define a margin around threshold voltage. As the margin value affects network performance, it should be chosen carefully. Similar to the computational model, the effect of previous synaptic weight on synaptic modification value is a very important factor which affects the network performance. 
The results of this study enhance our understanding of the requirements for implementing the reinforcement learning algorithm in hardware spiking neural networks. The findings of this study have a number of important implications for future practice on leveraging the required area on the device. The present study is limited to be used in large scale spiking neural networks because for representing  numbers, a large number of bits is currently required. Further research should be carried out to have a learning algorithms with lower required bits which are suitable for autonomous systems with large-scale spiking neural networks.  

\section{Conclusion}
This paper presents a digital multiplier-less architecture for an event-driven spiking neural network which is able to learn a context-dependent task through reinforcement learning. Neither neurons nor synapses use the high-cost functions like exponential or multipliers of original learning rules, therefore, our implementation leads to higher cost and area efficiency. 
As a proof of concept we implemented the proposed network on Xilinx Kintext-7 FPGA device. For validating the network performance on hardware, we connected the network to a robotic vehicle (Pushbot) in a closed sensors-network-motors loop. Using blinking LEDs that were located at different heights, a robot could distinguish among the different task-defining stimuli-triplets and successfully learn the rewarded associations between items and contexts, independent of the position of the items in the environment. The proposed model both on hardware with simulated inputs and in the robotic task reproduces the results of both the computational model and also to the animal experiment of previous studies~\cite{flor2014,Komo2009}. This work facilitates research to employ reinforcement learning for autonomous agents using hardware implemented SNN systems.
\section*{Acknowledgment}
The authors would like to thank Giacomo Indiveri, Dongchen Liang and Alessandro Aimar for fruitful discussions. We also thank Prof. J\"org Conradt for providing the robotic hardware and SNSF Project Ambizione (grant PZOOP2\_168183) for financial support.

\bibliographystyle{IEEEtran}
\bibliography{references}

\begin{IEEEbiography}
[{\includegraphics[width=1in,height=1.2in]{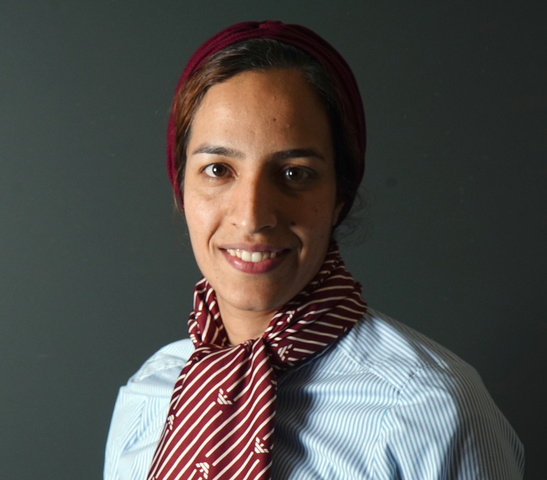}}]{Hajar Asgari}
received the M.Sc. degree in electrical engineering from Shahid Chamran University of Ahvaz, in 2013 and is a Ph.D. student in electrical engineering at Shahhid Beheshti University of Tehran, Iran. She is currently a visiting Ph.D. student with the Neuromorphic Cognitive System group, Institude of Neuroinformatic, university of Zurich (UZH) and ETH Zurich, Switzerland. Her current research focuses on neuromorphic engineering. 
\end{IEEEbiography}

\begin{IEEEbiography}[{\includegraphics[width=1in,height=1.2in]{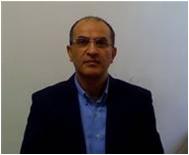}}]{Babak Mazloom-Nezhad Maybodi}
 received the B.Sc. and M.Sc. degree in Amirkabir University of Technology (Tehran Polytechnic) in 1985 and 1989, respectively. He received his PhD degree in September 2008 from University of Shahid Beheshti, Tehran, Iran. 
He was lecturer of Amirkabir University from 1985 to 1989. From 1989 until now, he is assistant professor of Shahid Beheshti University. Dr. M.-N. Maybodi research interests include Digital systems, Electronic Measurement and System Security. 

\end{IEEEbiography}

\begin{IEEEbiography}[{\includegraphics[width=1in,height=1.2in]{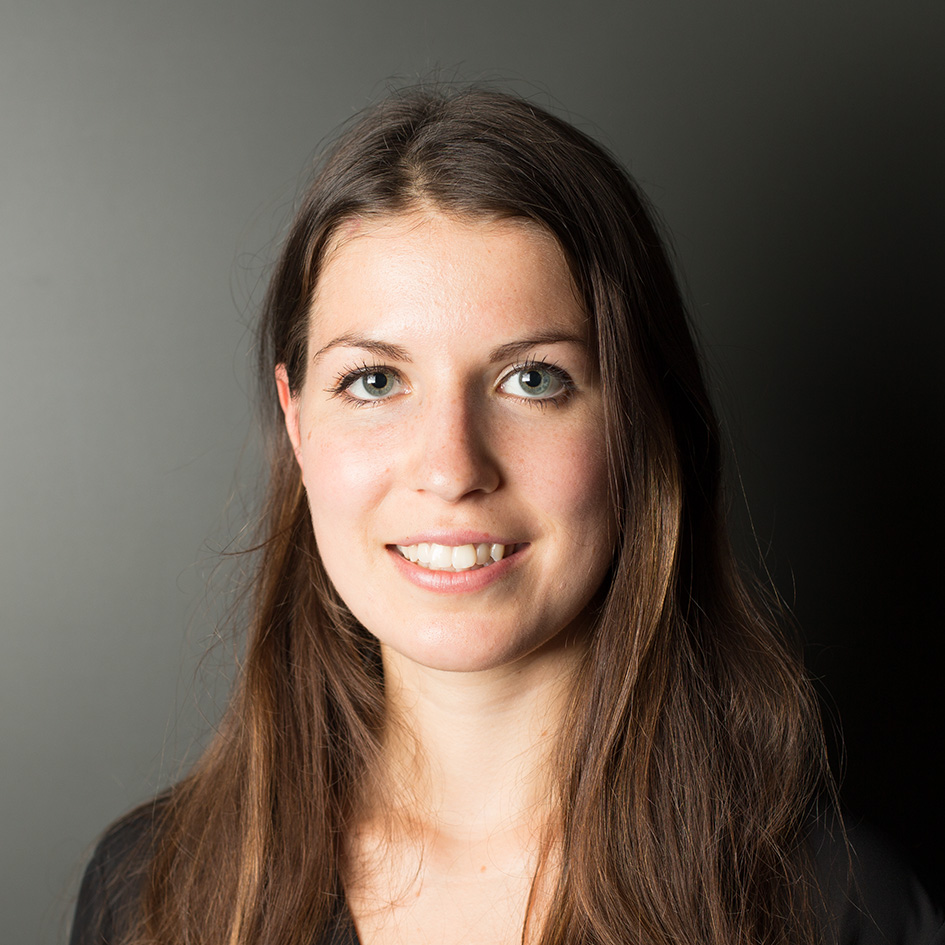}}]{Raphaela Kreiser} received the M.Sc. degree in "Neural Systems and Computation" at UZH and ETH Zurich. Currently she is a Ph.D. student in the Neuromorphic Cognitive Robots group at the Institute of Neuroinformatics, UZH and ETH Zurich, Switzerland.
Her research focuses on neuromorphic solutions for Simultaneous Localization and Mapping (SLAM).
\end{IEEEbiography}

\begin{IEEEbiography}[{\includegraphics[width=1in,height=1.1in]{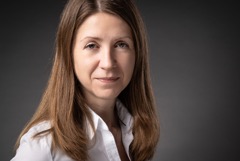}}]{Yulia Sandamirskaya}
is a group leader in the Institute of Neuroinformatics at the University of Zurich and ETH Zurich. Her group “Neuromorphic Cognitive Robots” studies movement control, memory formation, and learning in embodied neuronal systems and implements neuronal architectures in neuromorphic devices, interfaced to robotic sensors and motors. She has a degree in Physics from the Belorussian State University in Minsk and PhD from the Institute for Neural Computation in Bochum, Germany. She is the chair of EUCOG — European network for Artificial Cognitive Systems and coordinator of the NEUROTECH project (neurotechai.eu) that organises a community around neuromorphic computing technology.
\end{IEEEbiography}
\end{document}